\documentclass[useAMS,usenatbib,12pt,eps fig]{mn2e}
\usepackage[english]{babel}
\usepackage{amsmath,amssymb}
\usepackage{graphicx}
\usepackage{color}
\def\etal{et~al.}
\def\apj{ApJ}
\def\aj{AJ}
\def\mnras{MNRAS}
\def\pasj{PASJ}

\def\apjl{ApJL}

\def\be{\begin{equation}}
\def\ee{\end{equation}}
\def\bea{\begin{eqnarray}}
\def\eea{\end{eqnarray}}
\def\mum{\mu{\rm m}}

\title
[spatial extent and temperature of dust]
{Constraining spatial extent and temperature of dust around
galaxies from far-infrared image stacking analysis}
\author[Kashiwagi \& Suto]
{
\parbox[h]{\textwidth}{
Toshiya Kashiwagi$^{1}$,
Yasushi Suto$^{1,2,3}$}
\vspace*{2pt} \\
\hspace{-.1cm}$^1$Department of Physics, The University of Tokyo, 
Tokyo 113-0033, Japan \\
\hspace{-.1cm}$^2$Research Center for the Early Universe, School of
Science, The University of Tokyo, Tokyo 113-0033, Japan \\
\hspace{-.1cm}$^3$Department of Astrophysical Sciences, 
Princeton University, Princeton, NJ 08544, USA \\
}

\begin{document}

\date{Draft,\today}

\maketitle

\begin{abstract}
We propose a novel method to constrain the spatial extent of dust
around galaxies through the measurement of dust temperature. Our
method combines the dust emission of galaxies from far-infrared (FIR)
image stacking analysis and the quasar reddening due to the dust
absorption around galaxies.  As a specific application of our
method, we use the stacked FIR emission profiles of SDSS photometric
galaxies over the {\it IRAS} $100\mum$ map, and the recent measurement
of the SDSS galaxy-quasar cross-correlation.  If we adopt a
single-temperature dust model, the resulting temperature is around 18K,
which is consistent with a typical dust temperature for a central part
of galaxies.  If we assume an additional dust component with much
lower temperature, the current data imply the temperature of the
galactic dust needs to be higher, 20K to 30K. Since the model of
the density and temperature distribution of dust adopted in the current
paper is very simple, we cannot draw any strong conclusion at this
point. Nevertheless our novel method with the elaborated theoretical
model and multi-band measurement of dust will offer an interesting
constraint on the statistical nature of galactic dust.
\end{abstract}


\section{Introduction}\label{sec:introduction}

Dust plays important roles in cosmic star formation and evolution of the
galaxies.  The basic ingredients of dust grains are metals produced
through past stellar activity, and thus the main reservoir of dust is
conventionally thought to be mainly confined in interstellar space
within galaxies. \cite{Zwicky1962}, however, suggested the existence of
dust filling the intracluster space within the Coma cluster. This
motivated the investigation of the abundance and spatial distribution of
dust in different environments, including the color-excess of background
objects due to dust optical-UV reddening \citep{Zaritsky1994,
Chelouche2007, McGee2010, Muller2008}, and the FIR dust emission from
individual objects \citep{Stickel1998, Stickel2002, Kaneda2009,
Kitayama2009}, and from stacking analysis
\citep{Montier2005,Gutierrez2014}.

Recently, M{\'e}nard et al. (2010a: hereafter MSFR) investigated the
distribution of dust around galaxies by measuring the angular
correlation between SDSS galaxy distribution and distant quasar colors.
They found that the mean $g$-$i$ reddening profile around SDSS galaxies
is well approximated by a single power-law:
\begin{equation}
\langle E_{g-i}\rangle (\theta) = (1.5\pm0.4) \times 10^{-3}
 \left(\frac{\theta}{1'.0}\right)^{-0.86\pm0.19} ,
\label{eq:EgiMSFR}
\end{equation}
where $\theta$ is the angular separation between foreground galaxies and
background quasars. Furthermore they discovered that the above power-law
extends even for $\theta>10'$. The angular scale corresponds to several
Mpc at the mean redshift $\langle z\rangle =0.36$ of their SDSS galaxy
sample.  This is far beyond the typical scale of galactic disks, and
even larger than the virial radius of typical galaxy clusters.

MSFR appear to interpret their result as an evidence for an
extended dust surrounding an individual galaxy beyond a few Mpc, which
we refer to as {\it the circum-galactic dust model} (CGD
model). Their interpretation, however, is rather subtle. The mean
reddening profile from their measurement $\propto\theta^{-0.8}$ is close
to that of the angular correlation function of galaxies. Thus the
detected dust reddening may be equally explained by the summation of the
dust component associated with the central part of galaxies according to
the spatial clustering of those galaxies, which will be referred to as
{\it the inter-stellar dust model} (ISD model).

In practice, it is difficult to distinguish between the CGD and ISD
models on the basis of the statistical correlation analysis alone
as performed by MSFR. Therefore a complementary and independent method
to constrain the nature of the dust is needed. This is exactly what we
attempt to propose in this paper.

For that purpose, we measure the dust far-infrared (FIR) emission of the
SDSS galaxies by image stacking analysis.  Similar analysis on the SFD
Galactic extinction map \citep[SFD]{SFD1998} has detected the FIR
emission of SDSS galaxies \citep[KYS13]{Kashiwagi2013}.  We return to
the $100\mum$ intensity map by SFD, instead of their extinction map, and
perform the stacking analysis of the same galaxy sample used by MSFR.
If the detected FIR emission originates from the same dust component as
the MSFR reddening measurement, the emission to absorption ratio puts a
constraint on dust temperature, which would in turn offer complementary
information to distinguish between the CGD and ISD models
mentioned above.

The present paper is organized as follows. The data used in the current
analysis are described in Section 2.  In Section 3, we perform the
stacking analysis of the MSFR galaxy sample on IRAS/SFD $100\mum$ map.
We show the constraint on the dust temperature from the detected FIR
emission combined with the MSFR reddening measurement.  We present
summary and conclusions of the paper, and discuss future outlook in
Section 4.  Throughout the analysis, we assume the standard $\Lambda$CDM
cosmology with $\Omega_{\rm m}=0.3$, $\Omega_{\rm \Lambda}=0.7$, and $h
= 0.7$.

\section{Data}

We select our galaxy sample from the SDSS DR7 photometric galaxies with
5 passbands, $u$, $g$, $r$, $i$, and $z$, in northern galactic cap,
which covers $\sim$7600 ${\rm deg}^2$.  For details of the photometric
data, see \cite{Stoughton2002, Gunn1998, Gunn2006, Fukugita1996,
Hogg2001, Ivezic2004, Smith2002, Tucker2006, Padmanabhan2008, Pier2003}.
We conservatively masked $\sim 5\%$ of the total area following the SDSS
mask definition.  We also removed the objects with bad photometry or
fast-moving flag according to the photometry flags, which are suspicious
to be correlated with the Galactic foreground.  See \cite{Yahata2007}
for more details of our data selection.

For the current analysis, we impose the same $i$-band magnitude cut,
$17<m_i<21$, as the MSFR sample for a direct comparison with their
results, where the magnitudes of the galaxies are correct for Galactic
extinction using the SFD map \citep{SFD1998}. Our final sample collects
$2.88\times 10^7$ galaxies.

For far-infrared data, we use the all-sky diffuse 100$\mum$ map provided
by SFD; they have carefully processed the original IRAS/ISSA 100$\mum$
sky map, removing the scan pattern of {\it IRAS}, correcting calibration
errors based on COBE/DIRBE data, and subtracting zodiacal dust emission
and bright point sources with $f_{60\mum} > 0.6\rm{Jy}$.

Hereafter, we adopt a Gaussian with $\sigma=3'.1$ for the point spread
function (PSF) of SFD/IRAS map, as measured by similar stacking analysis
by KYS13.

\section{Image stacking analysis of FIR emission from SDSS galaxies}

\subsection{Stacked radial profiles}\label{subsec:model}

Following the procedures of KYS13, we stack the SFD/IRAS 100$\mum$ map
over  $120'\times 120'$ squares centered on each SDSS galaxy. Each
image is randomly rotated around the center.  The resulting stacked
image shows clear circular signature of dust emission associated
with those galaxies (KYS13).

\begin{figure}
\begin{center}
  \includegraphics[width=\hsize]{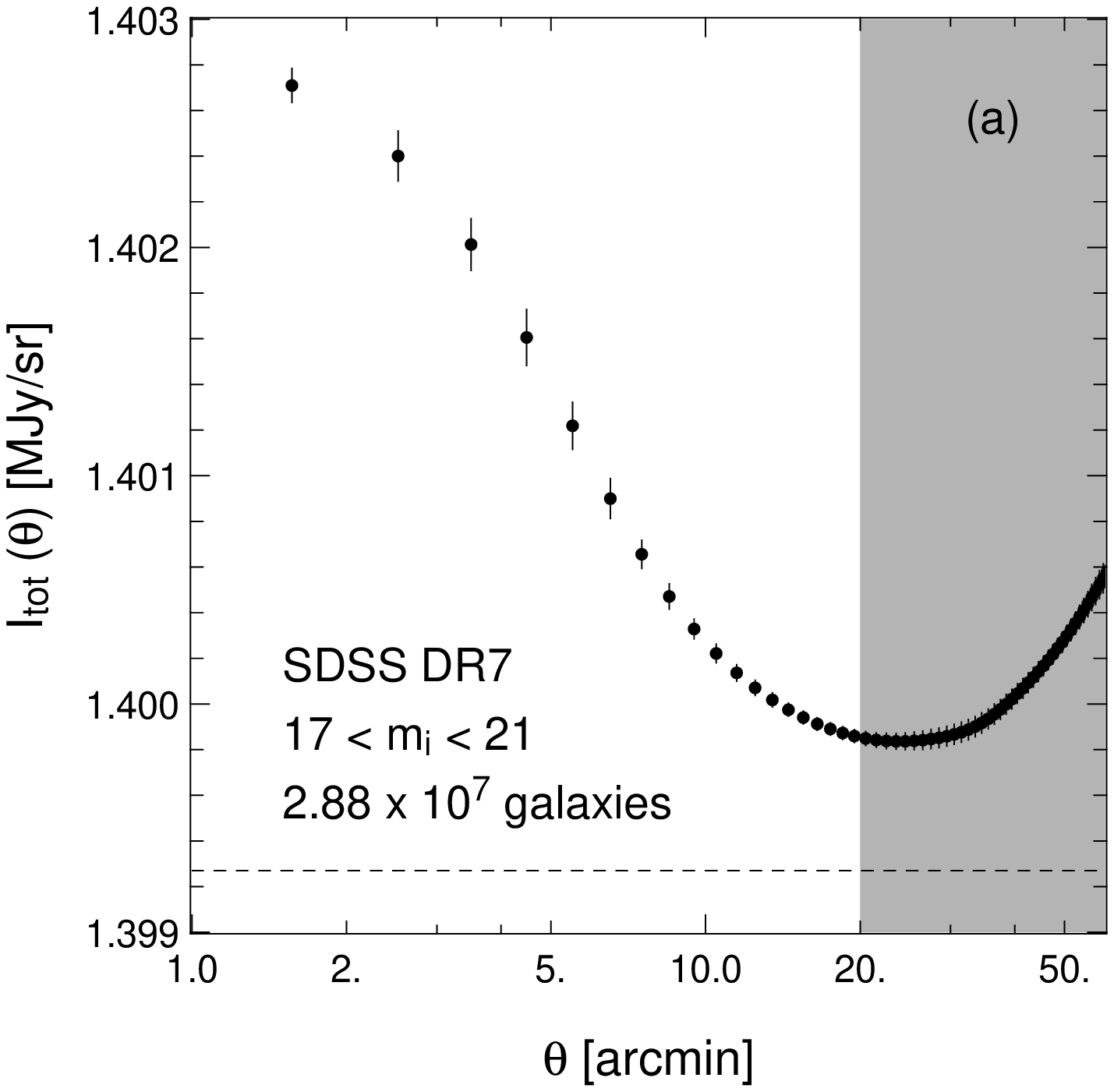}
  \hspace{5mm}

  \includegraphics[width=\hsize]{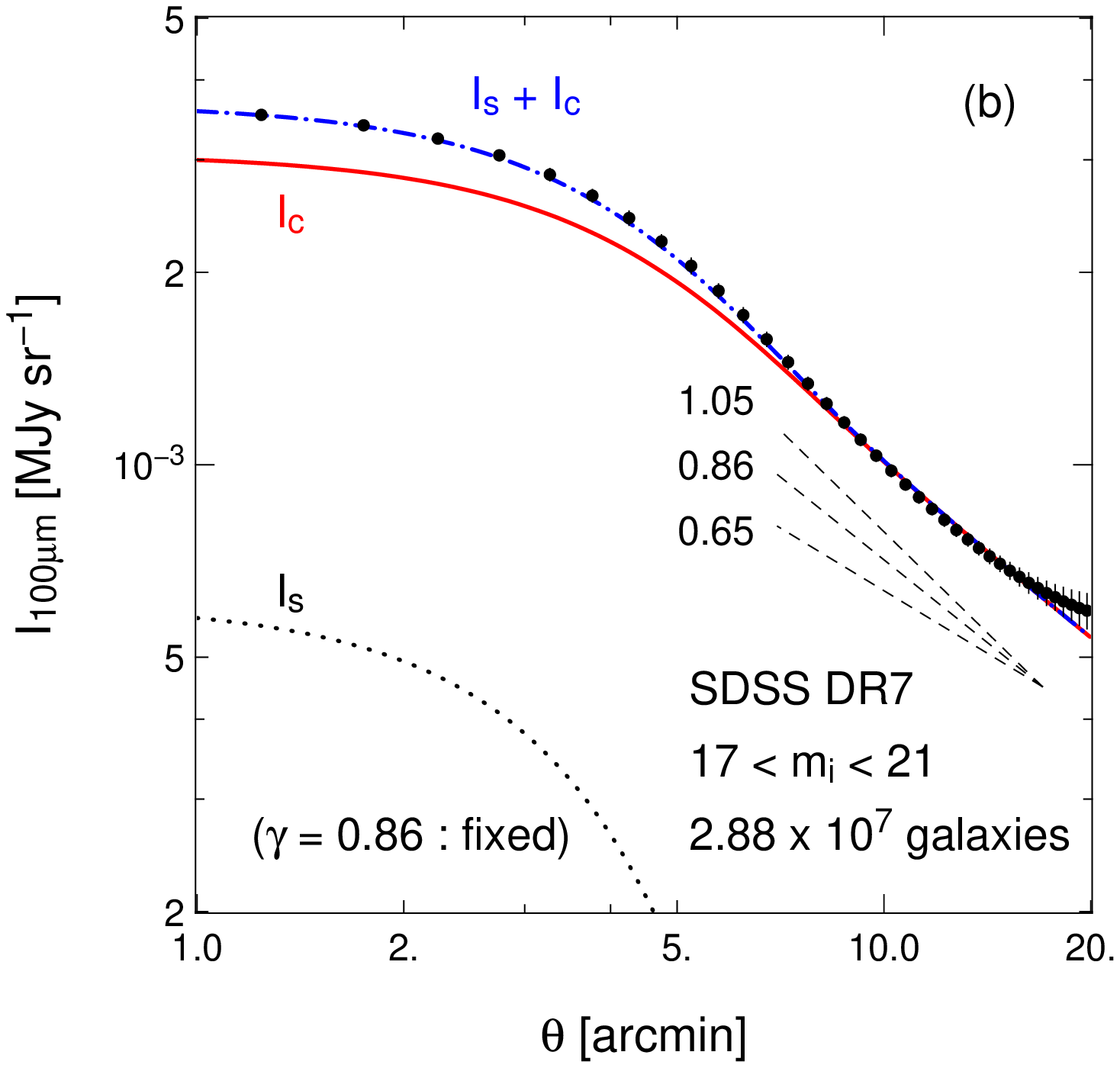} \caption{Radial profile of
the FIR stacked image of SDSS galaxies.  The symbols indicate the radial
average of the stacked image and the error bars show {\rm rms} in each
radial bin.  ({\it a}); Radial profile of the raw stacked image for
$\theta<60'$ (before subtracting the offset level due to the Galactic
dust).  Shaded region indicates the data with $\theta > 20'$ that are
not used in the current analysis.  ({\it b}); Radial profile of the
stacked image after subtracting the offset level of $C=1.39927
\rm{[MJy~sr^{-1}]}$, which is computed assuming $\gamma=0.86$ and shown
as the dashed line in panel a.  The lines indicate the best-fits
for $I_{\rm s}$ (black dotted), $I_{\rm c}$ (red solid), and $I_{\rm
tot}-C$ (blue dot-dashed); see equation (\ref{eq:model}).  Just for
reference, we plot the power-laws of $\gamma=0.65$, 0.86 and 1.05 in
dashed lines, which covers the ranges of the MSFR result in equation
(\ref{eq:EgiMSFR}). } \label{fig:profile}
\end{center}
\end{figure}

The radial profile of the raw stacked image is shown in Figure
\ref{fig:profile}a. The quoted error bars reflect the {\it rms} in each
radial bin ($\Delta\theta=1'.0$). The radial profile is reasonably
well fitted by Gaussian corresponding to the PSF around the
central region, but exhibits an extended tail beyond the PSF width,
$\sigma=3'.1$, which corresponds to roughly $1{\rm Mpc}$ for the mean
redshift $\langle z\rangle \sim0.36$ of the SDSS galaxies.

 At sufficiently large $\theta$, the stacked flux should be dominated by the 
Galactic foreground, which is uncorrelated with the SDSS galaxies and expected 
to be constant. The stacked flux, however, increases beyond $\theta>30'$. 
While we do not completely understand the behavior
(see also discussion in KYS13), it may be partly due to the fact that
the SDSS survey region is designed to be located at the low-extinction region, 
therefore towards high galactic latitudes. 
Thus the outskirt of the SDSS region is surrounded by low galactic latitudes with 
relatively higher values of the 100$\mum$ intensity, and the stacked flux centered at 
the SDSS region tends to be systematically larger at larger $\theta$. 
Nevertheless the profile for $\theta<20'$ 
matches nicely that expected from angular correlation functions of SDSS
galaxies (Okabe et al. in preparation). This is why we adopt the profile
modeling discussed below.
 
We adopt the following radial density profile of dust:
\begin{equation}
I_{\rm tot}(\theta) = I_{\rm s}(\theta) + I_{\rm c}(\theta) + C,
\label{eq:model}
\end{equation}
where $I_{\rm s}$ and $I_{\rm c}$ represent the contributions from the
central single galaxy (single term) and from the clustered neighbor
galaxies (clustering term), respectively, and $C$ is the background level of the foreground Galactic
dust emission \footnote{These
definitions of $I_{\rm tot}$, $I_{\rm s}$, and $I_{\rm c}$ are equivalent
to $\Sigma_{\rm g}^{\rm tot}$, $\Sigma_{\rm g}^{\rm s}$, and
$\Sigma_{\rm g}^{\rm c}$ used in KYS13, respectively,
except that $\Sigma_{\rm g}$ denotes the SFD map extinction in units of
[mag], whereas $I$ in this paper denotes the intensity in units of
[MJr/sr].}. 
We assume that the Galactic foreground, $C$, should be uncorrelated 
with the SDSS galaxies, and thus is assumed to be constant at $\theta<20'$.

Since the PSF of SFD/IRAS map is well approximated by Gaussian, 
$I_{\rm s}(\theta)$ is written as
\begin{equation}
\label{eq:Istheta}
I_{\rm s}(\theta) = I_{{\rm s}0} 
\exp \left(-\frac{\theta^2}{2\sigma^2}\right),
\end{equation}
where $\sigma=3'.1$ is the Gaussian width of PSF. 

The clustering term $I_{\rm c}$ is written in terms of $I_{\rm s}$ and
angular two-point correlation function (2PCF) of galaxy, $w(\theta)$, as
\begin{eqnarray}
I_{\rm c}(\theta) &=& \int dm' \frac{dN(m')}{dm} \nonumber \\
&~&~~~ \times \int d\mathbf{\varphi} 
I_{\rm s}(\mathbf{\theta}-\mathbf{\varphi};m') w(\mathbf{\varphi};m'),
\label{eq:cterm}
\end{eqnarray}
where $dN(m')/dm'$ is the differential number count of galaxies (whether
or not detected by SDSS) as a function of $m'$.  We assume that the
single term is written as a function of $m_i$ alone, therefore the
dependence on other physical quantities is neglected.  We approximate the
angular 2PCF is described as a single power-law in this angular scale
\citep{Connolly2002, Scranton2002};
\begin{equation}
\label{eq:wtheta}
w(\theta;m') = A(m')\left(\frac{\theta}{\theta_0}\right)^{-\gamma},
\end{equation}
where the amplitude $A$ is a function of $m_i$, but the index $\gamma$
is assumed to be a constant and independent of $m_i$.  In this case,
equation (\ref{eq:cterm}) reduces to
\begin{equation}
\label{eq:Ictheta}
I_{\rm c}(\theta) = I_{{\rm c}0}
\exp \left(-\frac{\theta^2}{2\sigma^2}\right) 
{~}_1F_1 \left(1-\frac{\gamma}{2};1;\frac{\theta^2}{2\sigma^2}\right),
\end{equation}
where ${}_1F_1(a;b;c)$ denotes the confluent hypergeometric
function, and
\begin{eqnarray}
\label{eq:Ic0}
I_{{\rm c}0} 
&=& 2\pi \sigma^2 \left(\frac{\varphi_0}{\sqrt{2}\sigma}\right)^{\gamma} 
\Gamma\left(1-\frac{\gamma}{2}\right) \nonumber \\
&~&~~~ \times \int dm' I_{{\rm s}0}(m') A(m')\frac{dN(m')}{dm'}.
\end{eqnarray}

We fit the radial profile of the stacked image using equations
(\ref{eq:model}), (\ref{eq:Istheta}), and (\ref{eq:Ictheta}).  In doing
so, we do not use equation (\ref{eq:Ic0}), but treat $I_{{\rm c}0}$
simply as one of the fitting parameters empirically determined from the
observed profile. Consistency of the resulting $I_{{\rm c}0}$ with
equations (\ref{eq:wtheta}) and (\ref{eq:Ic0}) independently measured
for SDSS galaxies is an interesting topic (KYS13), which will be
discussed in detail elsewhere (Okabe et al. in preparation).  We
estimate the statistical errors using the jackknife resampling method by
dividing the entire SDSS sky area into 400 patches of equal area.

 The detected emission profile at small $\theta$ is affected due to the
IRAS PSF, and should not be directly compared with the MSFR measurement.
Therefore, we use the clustering term, which is relevant for $\theta \gg
\sigma$, for the dust temperature constraint in the following Section.
In fact, the PSF effect on the clustering term vanishes at large
$\theta$ and equation (\ref{eq:Ictheta}) reduces to the power-law as
\begin{equation}
\label{eq:Icthetalim}
I_c(\theta) 
= \frac{I_{\rm c0}}{\Gamma\left(1-\gamma/2\right)}
\left(\frac{\theta}{\sqrt{2}\sigma}\right)^{-\gamma}. 
\end{equation}

Since we (implicitly) assume here that the mean reddening profile of
MSFR, equation (\ref{eq:EgiMSFR}), is explained in the clustered dust
model, the value of $\gamma$ in equation (\ref{eq:wtheta}) should match
the MSFR result.  In order to confirm the validity of the assumption, we
first choose $I_{{\rm s}0}$, $I_{{\rm c}0}$, $C$, and $\gamma$ as free
parameters, and fit to the observed profile imposing $I_{{\rm s}0}\geq0$
and $I_{{\rm c}0}\geq0$.  The resulting best-fit value, $\gamma = 1.07
\pm 0.16$, is consistent with that of MSFR, $\gamma=0.86 \pm 0.19$ (the
other best-fit values include $I_{\rm s0}= 0 {\rm[MJy~sr^{-1}]}$,
$I_{\rm c0}= (3.5 \pm 0.4) \times 10^{-3} {\rm[MJy~sr^{-1}]}$, and
$C=1.399\pm 0.035 {\rm [MJy~sr^{-1}]}$).  Indeed as Figure
\ref{fig:profile} illustrates, the difference among the predicted
profiles for $0.65<\gamma<1.05$ is very small for the angular scales of
our interest $\theta \gg \sigma$. The departure from the power-law for
$I_{\rm c} < 6\times10^{-4} {\rm [MJy~sr^{-1}]}$ is not a problem
because it simply reflects the sensitivity to the subtracted offset $C$.

Thus we fix $\gamma=0.86$ in what follows, and obtain the best-fit
parameters as $I_{\rm s0}= (6.1 \pm 4.0) \times 10^{-4}
{\rm[MJy~sr^{-1}]}$, $I_{\rm c0}= (3.1 \pm 0.7) \times
10^{-3}{\rm[MJy~sr^{-1}]}$, and $C=1.399\pm 0.035{\rm [MJy~sr^{-1}]}$.
The best-fit profile for each component is shown in Figure
\ref{fig:profile}.

The stacked FIR emission profile corresponding to the clustering term
for $\gamma=0.86$ is finally given as
\begin{equation}
\label{eq:stackprof}
\langle I_{100\mum}\rangle (\theta) 
= \frac{(7.0\pm1.6)\times 10^{-3}}{\rm{MJy~sr^{-1}}} 
\left(\frac{\theta}{1'.0}\right)^{-0.86},
\end{equation}
at large $\theta$, which plays a major role in our method proposed 
in Section \ref{subsec:constraints}.

We note that while the statistical error of $C$ is much larger than the
best-fit values of $I_{\rm s0}$ and $I_{\rm c0}$ themselves, it does not
affect the detection significance of the dust emission from SDSS
galaxies.  In fact, the variance of $C$ simply comes from that of the
Galactic dust over the SDSS survey area; the majority of the 400
jackknife subsamples indicates similar signatures of the dust emission,
except for the difference of $C$.

It is interesting to note that the observed stacked profile and the 
prediction from the summation of individual SDSS galaxies indeed 
agree very well, as mentioned in Section 1. Thus we will
proceed further to an independent and complementary analysis in order to
constrain the spatial extent of dust in the rest of this section.

\subsection{A simple model prediction of the dust emission} 
\label{subsec:model-prediction}

While the dust {\it extinction} is determined mainly by its column
density, the dust {\it emission} depends sensitively on its temperature
as well.  Therefore, if the measured extinction and emission comes from
the same dust distribution, their ratio serves as a sensitive measure of
the dust temperature. In this subsection, we will explicitly show
theoretical expressions for the reddening and emission of dust in a
simple model of dust density distribution.  Since we are interested in
the scales beyond the galactic disk scale, we consider the clustering
term alone.

The angular profiles of dust extinction and emission around a galaxy are
calculated by integrating the dust surface density $\Sigma_{\rm
d}(r_p,z)$ of nearby galaxies at $z$ separated by the projected distance
$r_p=d_A(z)\theta$ from the central galaxy, where $d_A(z)$ is the
angular diameter distance at $z$. For simplicity, we assume that the
2-dimensional projected dust surface density responsible for the
clustering term is given by a single power-law:
\begin{equation}
\label{eq:Sigma_d}
\Sigma_{\rm d}(r_p,z) = 
\Sigma_{{\rm d}0}(z)\left( \frac{r_p}{r_{p,0}} \right)^{-\gamma}.
\end{equation}
Throughout the current model, we set the power-law index $\gamma$ 
as that of the galaxy angular correlation function, equation
(\ref{eq:wtheta}), specifically $\gamma=0.86$ in what follows.  Although
we neglect the redshift evolution of the correlation length $r_{p,0}$,
it is effectively absorbed in $\Sigma_{{\rm d}0}(z)$ as
long as $\gamma$ is time-independent as assumed here.

Under the above assumptions, the angular extinction profile of dust at
 redshift $z$ is written as
\begin{eqnarray}
E_{g-i}(\theta,z) &=& 
\frac{2.5}{\ln 10}
\left[ \tau \left(\theta,\frac{\lambda_g}{1+z}\right) 
-\tau\left(\theta,\frac{\lambda_i}{1+z}\right) \right] \nonumber \\
&=& \frac{2.5}{\ln 10} 
\left[ \kappa_{\rm ext}\left(\frac{\lambda_g}{1+z}\right) - 
\kappa_{\rm ext}\left(\frac{\lambda_i}{1+z}\right) \right] \nonumber \\ 
&\times& \Sigma_{\rm d}\big(d_A(z)\theta,z\big),
\end{eqnarray}
where $\lambda_g$ and $\lambda_i$ are the rest-frame wavelengths of SDSS
$g$ and $i$-bands, respectively, and $\kappa_{\rm ext}(\lambda)$ is the
extinction cross-section per unit dust mass at a wavelength of
$\lambda$.  The average angular extinction profile around SDSS galaxies
is then given by
\begin{eqnarray}
\label{eq:extprof}
\langle E_{g-i} \rangle(\theta) 
&=& \frac{2.5}{\ln 10}
\left[\int^{\infty}_0 \frac{dN}{dz}dz\right]^{-1} \cr
&\times&
\int^{\infty}_0 \left[ \kappa_{\rm ext}
\left(\frac{\lambda_g}{1+z}\right) - \kappa_{\rm ext}
\left(\frac{\lambda_i}{1+z}\right) \right]  \cr
&~&\times~\Sigma_{\rm d0}(z)
\left(\frac{d_A(z)\theta}{r_{p,0}}\right)^{-\gamma}\frac{dN}{dz}dz,
\end{eqnarray}
where $dN/dz$ is the redshift distribution of SDSS galaxies. Following
MSFR, we adopt an approximation \citep{Dodelson2002}: 
\begin{equation}
\label{eq:dNdz}
\frac{dN}{dz} \propto z^2 e^{-(z/0.187)^{1.26}}.
\end{equation}
Thus the number-weighted mean redshift of the sample is given by
\begin{equation}
\langle z\rangle = \frac{\int z (dN/dz) dz}{\int (dN/dz)dz} = 0.36.
\end{equation}

One can similarly compute the angular FIR emission profile around SDSS
galaxies. Since the dust emission at $\lambda=100\mum$ is well
approximated by the blackbody spectrum, the corresponding surface
brightness at redshift $z$ is given as
\begin{align}
I_{100\mum}(\theta,z,T_{\rm d}) &=
\frac{1}{(1+z)^4}B_{\nu}\left(\frac{100\mum}{1+z},T_{\rm d}\right) \cr
& \times \kappa_{\rm abs}\left(\frac{{100\mum}}{1+z}\right)
\Sigma_{\rm d}\big(d_A(z)\theta,z\big),
\end{align}
where $\kappa_{\rm abs}$ is the absorption cross section per unit dust
mass, $B_{\nu}$ is the blackbody spectrum per unit frequency, $T_{\rm
d}$ is the dust temperature, which we assume to be independent of
$z$, and the same for all SDSS galaxies, and $1/(1+z)^4$ comes from the
cosmological dimming effect.

The average angular emission profile of SDSS galaxies, which corresponds
to $I_{\rm c}(\theta)$ observed by the stacking analysis, is given as
\begin{eqnarray}
\label{eq:FIRprof}
&& \langle I_{100\mum}\rangle (\theta,T_{\rm d})  \cr
&=& 
\left[\int^{\infty}_0 \frac{dN}{dz}dz\right]^{-1} ~
\int^{\infty}_0 \frac{1}{(1+z)^4}B_{\nu}
\left(\frac{{100\mum}}{1+z},T_{\rm d}\right) \cr
&\times& \kappa_{\rm abs}\left(\frac{{100\mum}}{1+z}\right) 
\Sigma_{\rm d0}(z)\left(\frac{d_A(z)\theta}{r_{p,0}}\right)^{-\gamma}
\frac{dN}{dz}dz .
\end{eqnarray}

Because we adopt the power-law dust profile, equation
(\ref{eq:Sigma_d}), the ratio of equation (\ref{eq:FIRprof}) to
(\ref{eq:extprof}) is independent of $\theta$, and written in terms of
$\kappa_{\rm ext}$, $\kappa_{\rm abs}$, and $T_{\rm d}$ alone.

\begin{figure}
\begin{center}
  \includegraphics[width=\hsize]{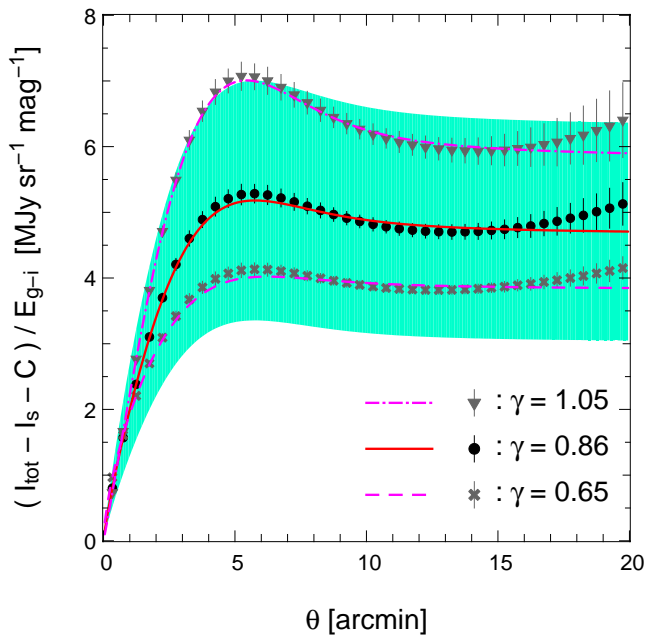}
\caption{Radial profile of $100\mum$ emission to reddening ratio.
Filled circles (black) indicate the observed $100\mum$ emission profile
of the stacking analysis divided by the MSFR reddening profile, where
the best-fit single term $I_{\rm s}(\theta)$ and the offset level $C$
assuming the 2PCF index as $\gamma = 0.86$ are subtracted.  Red solid
curve is the best-fit profile of the clustering term $I_{\rm c}(\theta)$
with $\gamma=0.86$, divided by $E_{g-i}(\theta)$ of MSFR.  Shaded region
indicates the uncertainty of the emission to reddening ratio where the
statistical error of $I_{\rm c0}$ and the MSFR measurement is taken into
account.  Crosses (filled triangles) and dashed (dot-dashed) curve
indicate the same as filled circles and solid curve, but for
$\gamma=0.65~(1.05)$. }  \label{fig:IFIR2Egi}
\end{center}
\end{figure}

The observed profile of the emission to reddening ratio is shown in
Figure \ref{fig:IFIR2Egi}.  The filled circles are plotted using the
residual of the emission profile, from which the best-fit single term
and the offset level assuming $\gamma = 0.86$ are subtracted.  The red
solid curve shows the ratio of the best-fit clustering term, $I_{\rm
c}(\theta)$, with $\gamma=0.86$ to equation (\ref{eq:EgiMSFR}), and the
shaded region indicates its uncertainty due to the statistical error of
$I_{\rm c0}$ and the amplitude of equation (\ref{eq:EgiMSFR}).  The
uncertainty of the power-law index in equation (\ref{eq:EgiMSFR}) is not
considered here.  At small $\theta$, the emission profile is suppressed
due to the SFD/IRAS PSF effect, whereas the ratio converges to a
constant at large scale.  The emission to reddening ratio at large
$\theta$ limits is given by equations (\ref{eq:EgiMSFR}) and
(\ref{eq:stackprof}) as $\langle I_{100\mum}\rangle/\langle
E_{g-i}\rangle =4.7\pm1.6~{\rm [MJy~sr^{-1}mag^{-1}]}$, which
corresponds to the shaded regions in Figure \ref{fig:Td} below.

We also consider the extent to which this result is sensitive to the
choice of the power-law index $\gamma$, which is fixed as 0.86 in the
analysis above.  We repeat both the fitting to the observed profile and
the theoretical calculation of equation (\ref{eq:extprof}) and
(\ref{eq:FIRprof}), varying the value of $\gamma$ from 0.65 to 1.05.
Figure \ref{fig:IFIR2Egi} shows the observed emission to reddening ratio
for $\gamma=0.65$, $0.86$ and $1.05$.  The average ratio changes
approximately $\sim20$ per cent (and its fractional uncertainty is
similar to that for the case of $\gamma=0.86$, although it is not shown
in Figure \ref{fig:IFIR2Egi}).  We also make sure that the theoretical
value from equation (\ref{eq:extprof}) and (\ref{eq:FIRprof}) changes by
$10$ per cent according to the corresponding change of $\gamma$.
Consequently, we find that the uncertainty of dust temperature due
to the choice of $\gamma$ is merely $\sim 1$K. This can be neglected
comparing with the possible larger systematics due to other many
simplifying assumptions.

\subsection{Constraints on dust temperature}\label{subsec:constraints}

\begin{figure}
\begin{center}
  \includegraphics[width=\hsize]{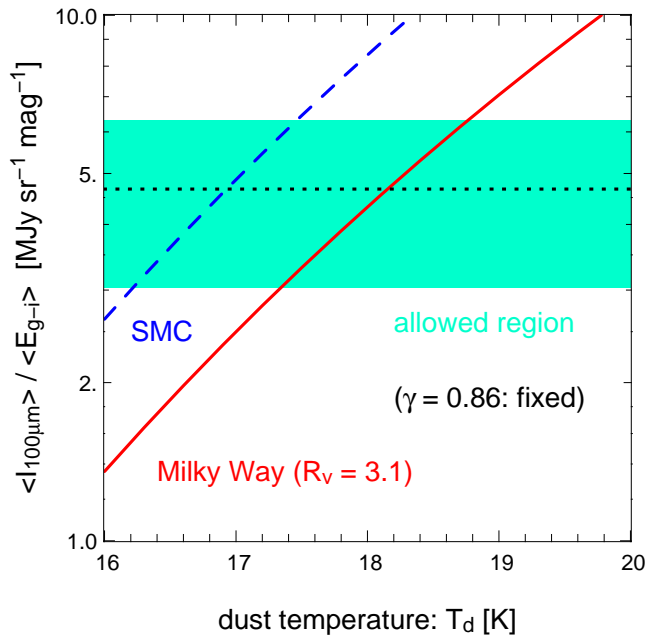} 
\caption{Constraints on the FIR emission to extinction ratio from MSFR
  and the stacking results (shaded region).  Solid and dashed lines
  indicate the prediction for Milky Way ($R_V=3.1$) and SMC dust model
  \citep{Weingartner2001}, respectively.  The power-law index of galaxy
  2PCF is fixed as $\gamma=0.86$.  If $\gamma=0.65$ and $1.05$ is
  assumed, the allowed region is shifted by $-20$ and $20$ per cent,
  respectively.} \label{fig:Td}
\end{center}
\end{figure}

The solid and dashed lines in Figure \ref{fig:Td} indicate the expected
emission to extinction ratio as a function of $T_{\rm d}$.  We adopt the
values of $\kappa_{\rm ext}$ and $\kappa_{\rm abs}$, from the dust model
by \cite{Weingartner2001} \footnote{Data is taken from Web-site of
B. T. Draine, http://www.astro.princeton.edu/~draine/dust/dustmix.html.}
for Milky Way ($R_V=3.1$) and SMC dust, for solid and dashed lines,
respectively.

Here the redshift dependence of $\Sigma_{\rm d0}(z)$ is neglected and
assumed to be constant just for simplicity.  Indeed we made sure
that the $z$-dependence of $\Sigma_{\rm d0}(z)$ does not significantly
change the result; if we assume $\Sigma_{\rm d0}(z) \propto (1+z)^p$ for
instance, the model prediction of $\langle I_{100\mum}\rangle/\langle
E_{g-i}\rangle$ changes by $\mp 15$ per cent for $p=\pm 1$, and the dust
temperature constraint changes by $\pm 0.2{\rm K}$.  The recent
measurement of dust mass function by \citet{Dunne2011} found that the
cosmic dust mass density in sub-mm galaxies rapidly increases with
redshift up to $z\sim0.5$.  Thus the constraint on the dust temperature
below may be slightly underestimated.

Figure \ref{fig:Td} indicates that the dust model predictions and the
observed region are consistent if $T_{\rm d}=18.2^{+0.6}_{-0.9}{\rm K}$
for MW dust, and $T_{\rm d}=17.0^{+0.5}_{-0.8}{\rm K}$ for SMC, thus the
obtained constraints are almost insensitive to the choice of dust model.

Given several approximations adopted in our simple model, the quoted
statistical errors may underestimate the real uncertainty of the dust
temperature.  Nevertheless it is encouraging that the derived dust
temperature is in good agreement with that of the typical cold component of 
the ISD \citep{SFD1998, Dunne2011, Clemens2013}.
While this result is reasonably consistent with the ISD model, it is
premature to conclude that the observed dust profile is explained by the
sum of the dust in the central parts of individual galaxies. Our dust
model above is based on a single-temperature component, and the
extended dust component around a galaxy may have a substantially lower
temperature. In this case, the emission flux is much smaller while the
reddening amplitude remains almost the same.

Therefore we consider a two-component dust model below. In order to
keep the same surface density profile of dust, we assume that the two
components, corresponding to ISD and CGD, share the identical spatial
distribution, but they are allowed to have different temperatures
$T_{\rm ISD}$ and $T_{\rm CGD}$.  While this is still a very simple
model, we would like to proceed with it because the low-angular
resolution of IRAS images make it difficult to distingush the density
profile less than $\sigma=3'.1$. In any case the purpose of the current
paper is to propose a new method to constrain the dust density and
temperature, which will be improved significantly later theoretically
and observationally.

In the above spirit, we replace equation (\ref{eq:Sigma_d}) by
\begin{equation}
\label{eq:Sigma_d_CGM}
\Sigma_{\rm d}(r_p,z) = 
[\Sigma_{{\rm d}0,{\rm ISD}}(z)+ \Sigma_{{\rm d}0,{\rm CGD}}(z)]
\left( \frac{r_p}{r_{p,0}} \right)^{-\gamma}.
\end{equation}
 Then the observed FIR emission to extinction ratio becomes
\begin{align}
\label{eq:ratio}
& \hspace*{-0.5cm} 
\bigg[\frac{ \langle I_{100\mum} \rangle}
{\langle E_{g-i}\rangle} \bigg]_{\rm obs}= \cr
& \hspace*{-0.5cm} 
~\frac{f_{\rm ISD} \langle I_{100\mum} \rangle(T_{\rm ISD}) 
+ (1-f_{\rm ISD}) \langle I_{100\mum}\rangle (T_{\rm CGD})}
{ \langle E_{g-i} \rangle }, 
\end{align}
 at large $\theta$, where $\langle E_{g-i}\rangle$ and $\langle
I_{100\mum}\rangle(T)$ in the right-hand-side are given by equation
(\ref{eq:extprof}) and (\ref{eq:FIRprof}), respectively. We further
assume that the fraction of the ISD mass to the total mass:
\begin{equation}
\label{eq:ratio_ISM}
f_{\rm ISD} = \frac{\Sigma_{{\rm d}0,{\rm ISD}}}
{ \Sigma_{{\rm d}0,{\rm ISD}} + \Sigma_{{\rm d}0,{\rm CGD}} }
\end{equation}
is independent of redshift. Thus the observed value of $\langle
I_{100\mum} \rangle/\langle E_{g-i}\rangle$ provides a constraint on
$f_{\rm ISD}$ for given values of $T_{\rm ISD}$ and $T_{\rm
CGD}$. The preceding analysis corresponds to $f_{\rm ISD}=1$. 

 The temperature of the CGD is fairly uncertain because of the
unknown heating mechanism of the CGD.  If the heating source of the CGD
is dominated by the cosmic UV background, which is lower than the
interstellar radiation field in the solar neighborhood by two orders of
magnitude \citep{Madau2000,Gardner2000,Xu2005}, we obtain $T_{\rm
CGD}=10{\rm K}$ following \citep{Draine1984,Draine2011}.  On the other
hand, \citet{Yamada2005} assume the collisional heating mechanism by hot
plasma and the efficient injection of dust grains outside the galactic
disk. In this case, they suggested a possibility that the dust
temperature reaches even $\sim30{\rm K}$.

 The abundance of such possible high-temparature CGD, however, is
severly constrained by the observed data. \citep{Clark2015} fitted the
SED of dust-selected galaxies, and found that the majority of them have
cool and warm dust components with $T\sim (10$--$20)$K and $> 30$K. The
mass of the cool component is typically 100 times larger than that of
the warm component.  While the two components may correspond to the
two-temperature phases in ISD, the estimated mass ratio can be also
interpreted to put a severe constraint on the presence of the hot CGD
with $T\sim 30$K.  Furthermore, \cite{Draine2014} reported that the dust
temperature near the edge of M31 disk is $15{\rm K}$. Thus the dust
temperature in the outskirt is naturally expected to be much lower. 

For those reasons, we adopt $T_{\rm CGD}=10{\rm K}$ in what
follows.  We also confirm that the result below does not change as long
as $T_{\rm CGD}$ is lower than $10$K.

\begin{figure}
\begin{center}
  \includegraphics[width=\hsize]{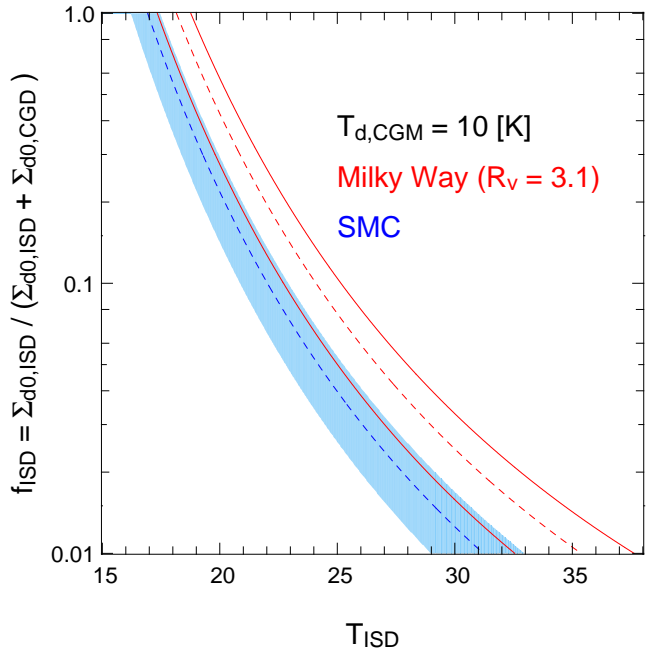} \caption{Constraints
on the ISD temperature, $T_{\rm ISD}$, and the mass fraction of the ISD
to the sum of the ISD and CGD.  We assume that the CGD temperature 
as $T_{\rm CGD}=10~{\rm K}$. Red curves and blue shaded
region indicate the constraint adopting Milky Way ($R_V=3.1$) and SMC
dust model \citep{Weingartner2001}, respectively.  }
\label{fig:fISM-Td}
\end{center}
\end{figure}

 Figure \ref{fig:fISM-Td} shows the constraint on $T_{\rm ISD}$ and
$f_{\rm ISD}$ from the observed value of $\langle I_{100\mum}
\rangle/\langle E_{g-i}\rangle$.  Due to the strong degeneracy between
the two parameters, the constraint allows a wide range of the ISD mass
fraction, even as small as $f_{\rm ISD}\sim 10^{-2}$ if $T_{\rm ISD}
=30{\rm K}$.  Thus the measurement of the mean dust temperature of the
central parts of galaxies $T_{\rm ISD}$ is crucial in distinguishing the
origin of the spatial extension of dust.  Indeed, the average
temperature of the ISD varies depending on the properties of the
galaxies, and can be as high as $\sim40{\rm K}$ \citep{Skibba2011}.  In
this respect, the current data do not exclude a possibility that a
substantial amount of the CGD exists, as suggested by MSFR and the
subsequent studies \citep[{e.g.,}][]{Fukugita2011, Menard2012,Peek2014}.
Nevertheless further improvements in model predictions and the
observations in future would put more stringent constraints on the
spatial extent of dust through the measurement of dust temperature as we
proposed in this paper.

\section{Summary \& Conclusions}

The spatial distribution of dust is of fundamental importance in
understanding the star formation and metal circulation history in the
universe. It is also crucial in correcting for the magnitude of distant
objects due to the resulting reddening/extinction \citep{Aguirre1999,
Menard2010,Fang2011}.

In a previous paper \citep{Kashiwagi2013}, we have detected the FIR dust
emission from SDSS galaxies via their image stacking analysis. We found
that the amount of dust emission is largely responsible for the observed
anomaly in the surface density of SDSS galaxies as a function of the SFD
extinction \citep{Yahata2007, Kashiwagi2015}.  Our previous analysis
implicitly assumed that the dust of each galaxy is locally confined in
the galactic disk scale, and that the observed FIR emission within the
large PSF width (FWHM$=6'.1$) is simply given as a sum of contributions
of individual galaxies (corresponding to the ISD model in the
present paper).  In contrast, the dust around a galaxy may be indeed
spatially extended up to $\sim 1$Mpc (CGD model), as claimed by
\citet{MSFR2010} and more recently by \citet{Peek2014} through the
correlation of background object colors against the separation length of
foreground galaxies.

In order to distinguish between the ISD and CGD models, we propose
a new method that constrains the temperature of dust by combining the
absorption (detected through reddening of quasars) and emission
(detected through the stacking of galaxies) features. Assuming that the
nature of galactic dust is described by those of MW and SMC, we find
that the observed dust is reasonably explained in terms of ISD model if
the dust temperature of the central parts of individual galaxies,
$T_{\rm ISD}$, is approximately 20K. The estimated temperature is
consistent with that of the galactic dust in the central region, but may
be higher than that predicted for CGD, if it is heated by UV
background alone. On the other hand, the substantial amount of
dust may reside far outside the galactic disks if $T_{\rm ISD}$ is much
higher than 20K.

Given several simplification and approximations that we adopted in the
present simple model analysis, the associated error-bars of the derived
dust temperature is fairly uncertain. Nevertheless we would like to
emphasize that the main purpose of the present paper is to propose a new
observational method to diagnose the nature of galactic dust.  Therefore
we do not discuss the interpretation of the present preliminary result.

Our proposed method should be, and indeed can be, improved in many
ways; the two components of dust may have different spatial density
profiles in addition to the different temperatures.  The redshift
evolution of the temperature and amount of dust may be included in the
theoretical models.  In addition, the line-of-sights of our emission and
reddening measurements may not be exactly the same. Since quasars behind
the heavily extincted line-of-sights may not be identified in the SDSS
photometric catalogue, the reddening measurement may systematically
underestimate the real optical depths while the emission is largely free
from the bias. Those improvements of the theoretical models and the
effect of the possible selection bias need to be investigated, for instance, 
with mock simulations, which is beyond the scope of the present paper.

The observational data and analysis can be also improved in future.
The dust temperature would vary depending on the different properties of
galaxies, and the amount of dust emission should depend on the
morphology of galaxies.  The image stacking analysis with better angular
resolutions and in multi-wavelengths would significantly improve the
observational data. Indeed current result is significantly limited by
the poor angular resolution of {\it IRAS}.  In those respects, the
higher-angular-resolution and multi-band far-infrared data by {\it
AKARI} \citep{Murakami2007} are very promising. We plan to present
elsewhere more detailed and systematic results using the {\it AKARI}
data (Okabe et al. in preparation).

\bigskip 

We thank Brice M{\'e}nard, Bruce T. Draine, Masato Shirasaki, and Tetsu
Kitayama for useful discussions. We also appreciate several
comments by an anonymous referee that motivated us to consider the
two-temperature dust model as well. This work is supported in part from
the Grant-in-Aid No. 20340041 by the Japan Society for the Promotion of
Science.  Y.S. and T.K. gratefully acknowledge supports from the Global
Scholars Program of Princeton University and from Global Center for
Excellence for Physical Science Frontier at the University of Tokyo,
respectively.

Data analysis were in part carried out on common use data analysis
computer system at the Astronomy Data Center, ADC, of the National
Astronomical Observatory of Japan.

Funding for the SDSS and SDSS-II has been provided by the Alfred P.
Sloan Foundation, the Participating Institutions, the National Science
Foundation, the U.S. Department of Energy, the National Aeronautics and
Space Administration, the Japanese Monbukagakusho, the Max Planck
Society, and the Higher Education Funding Council for England.  The SDSS
Web Site is http://www.sdss.org/.

The SDSS and SDSS-II are managed by the Astrophysical Research
Consortium for the Participating Institutions. The Participating
Institutions are the American Museum of Natural History, Astrophysical
Institute Potsdam, University of Basel, Cambridge University, Case
Western Reserve University, University of Chicago, Drexel University,
Fermilab, the Institute for Advanced Study, the Japan Participation
Group, Johns Hopkins University, the Joint Institute for Nuclear
Astrophysics, the Kavli Institute for Particle Astrophysics and
Cosmology, the Korean Scientist Group, the Chinese Academy of Sciences
(LAMOST), Los Alamos National Laboratory, the Max-Planck-Institute for
Astronomy (MPIA), the Max-Planck-Institute for Astrophysics (MPA), New
Mexico State University, Ohio State University, University of
Pittsburgh, University of Portsmouth, Princeton University, the United
States Naval Observatory, and the University of Washington.


\end{document}